\documentclass[12pt]{iopart}
\usepackage{iopams,graphics,graphicx,epstopdf,color,comment} 
\begin{document}
\title{A self-organized graph evolution model with preferential network random walk}
\author{S. Mehraban}
\address{Department of Physics, Sharif University of Technology, Tehran P.O. Box 11155-9161, Iran}
\ead{saeed.mehraban90@gmail.com}
\author{M.R. Ejtehadi}
\address{Department of Physics, Sharif University of Technology, Tehran P.O. Box 11155-9161, Iran}
\ead{Ejtehadi@sharif.edu}
%%%%%%%%%%%%%%%%%%%%%%%%%%%%%%%%%%%%%%%%%%%%%%
\begin{abstract}
We introduce a self-organized model of graph evolution associated with preferential network random walkers. The idea is developed by using two different types of walkers, the interactions of which lead to a dynamic graph. The walkers of the first type cause an enhancement in link attachments, while the second types have a destructive behavior. The statistical properties of the resulting network, including weight distributions, clustering, spectral densities and average path length are evaluated. As the ratio of the population of two types is balanced, the network faces a phase transition. We show that in the transition point, the graph behaves as a scale-free network, with a scaling exponent of $\sim -1.7$. 
\end{abstract}
\maketitle
%%%%%%%%%%%%%%%%%%%%%%%%%%%%%%%%%%%%%%%%%%%%%%%%%%%%%%%%%%
\section{Introduction}
Complex systems in nature and technology, ranging from neural networks to the World Wide Web, are well described by complex network theory~\cite{Barab2002, Newman2001, Newman2003}. There have been some efforts to describe the behavior of several real-world networks, such as cells, nervous systems, social organizations, etc., by means of graph theory~\cite{Barab2000,Ghall99,Newman2002-1}. The elements of such systems are represented as nodes of a graph, while the interactions among the elements are modeled by the graph's edges. The strengths of the interactions are considered to be the edges' weights. For example, a cell ~\cite{Weng99} is described by its chemicals. The interactions among these chemicals are the chemical reactions. In case of a nervous system ~\cite{ Koch2006}, the nodes are nerve cells, connected by axons. In social networks ~\cite{Wasserman}, the organizations and their connections can also be modeled by nodes and edges of a graph, respectively.

A random graph is an ensemble of $N$ nodes that are randomly attached together. Each two nodes in the ensemble are connected together with a certain probability. It is observed that as the number of nodes in a random graph is increased, the degree distribution takes a Poisson~\cite{Erdos59,Erdos60}. Although random graphs are perfect solvable models, because of wrong clustering, and also unrealistic Poisson degree distribution, they deviate from real network systems~\cite{Newman2002}. Recent studies have revealed that many real-world networks behave as scale-free. The evolution of these networks has roots in two mechanisms; the network growth and preferential attachment~\cite{Barab2002,Barab2000,Barab2000-1}. For example, in the World Wide Web it is observed that the more linked the elements are, the more they prefer to have links to other elements ~\cite{Barab99-1}. This behavior is called the preferential attachment, which is not included in random graph theory.

The construction of a graph is concerned with the algorithm of network growth. Depending on the structure of this algorithm, the resulting network can be categorized into several topological groups ~\cite{Bocc2006}, including Erd{\H o}s network, scale-free networks and small-world networks. In an Erd{\H o}s random graph certain numbers of links are selected randomly among all potential edges. The degree distribution ${\it P(k_i=k)}$ is the probability that the node $i$ has $k$ links. For different nodes $i$ and $j$, $P(k=k_i)$ and $P(k=k_j)$ are close to being independent random variables. Hence, it suffices to determine the number of nodes with degree $k$, $X_k$, and then $P(k)=X_k/\sum_{k} X_k$. For random graphs, the degree distribution is a binomial, and if the number of nodes $N$ is large, the distribution is Poisson ~\cite{Barab2002}.

In comparison to an Erd{\H o}s random graph, in which attachment of each particular node is uncorrelated to its degree, a scale-free network is concerned with preferential attachment. Recent studies ~\cite{Barab99}, regarding construction of a scale-free network, reveal that the degree distribution in such networks is shaped with a power-law tail.
The measure characterizing the neighborhood structure of the network is the clustering coefficient, i.e., the number of triangles, which are connected to one vertex ~\cite{Barrat2000}. Being inspired by clustering, we can discuss a class of networks, that although the most nodes of which are not neighbors, average path lengths between the nodes are short. This type of network is called small-world. The average distance between nodes for a small-word topology, after a cross-over size, grows logarithmically as the network's size grows~\cite{Watts99, Barab2002, Newman00}.

In this paper, we propose a model of a dynamic graph, in which the dynamics is due to the mobility of random walkers. In order to have the network evolve, we used two types of walkers; the walkers of the first type (constructive walkers) tend to enhance the transition probabilities of the links they have just walked through, while the walkers of the second type (destructive walkers) have a decreasing effect. We examined the statistics of link attachment probabilities (transition rate) and the popularity of the edges (traffic rate), as the graph evolves. The results reveal that although local variables critically depend on initial conditions, global parameters are self-organized and completely independent of initial conditions. Depending on the ratio of the walkers' populations (number of constructive walkers to the number of destructive walkers) due to a phase transition, the evolved network in its steady state is categorized in three states. We show how scaling in weight distributions emerge, as the network reaches to its critical point of transition.

This modeling can be applied to several different physical systems concerning dynamic interactions among elements. There are several real systems in which elements have similar properties as those of constructive walkers. For example, the dynamics of synaptic weights which are assumed to be the principal mechanism of memory storage in nervous systems, are believed to obey Hebb's law,``the cells which fire together will tend to wire together'' ~\cite{Rossum2000,Heerama99}. Transportation systems can also be modeled by this dynamic random walk, since the popular pathways in such systems tend to absorb more populations~\cite{Barrat2004}. Moreover, the effects of destructive walkers struggling with constructive ones, are realizable in a variety of applications. Using this modeling one can study the propagation of two different types of substances through a network of pipelines. The first type of substance tends to sediment through the pipe, while the second type tends to facilitate the propagation of water.

%The rest of this paper is organized as follows: after a review on related graph models, we introduce our model. Analytical %evaluation of the problem is indicated in \emph{part II.I}. The paper proceeds analyzing the order and the disorder in the %network, using the \emph{average Shannon entropy} and several \emph{order parameters}. The main focus of this work %is to study the statistics of the network in its steady state. In part \emph{section III} after estimating the relaxation time, %weight distributions are demonstrated. Afterwards, the spectral properties of Network in different states are %investigated. In order to recognize further topological features of the network, \emph{Part III.III} is dedicated to the %study of average distances between nodes, and average clustering coefficient in different graph states.
%%%%%%%%%%%%%%%%%%%%%%%%%%%%%%%%%%%%%%%%%%%%%%%%%%%%%%%%%%%%%%%%%
\section{The model}\label{The model}
The model introduced in this paper is based on random walk of walkers on an interactive dynamic graph ~\cite{Yang2005}. We start with a fully connected network of $N$ nodes. We put some random walkers on the network. At any time-step, each walker chooses one of its neighbors with a certain probability and transits to the chosen neighbor. The walkers are considered to be of two classes. The first class is called constructive (CW), and the second class is called destructive (DW). The former tends to enhance the transition probability of the link it has just walked through, while the latter decreases this probability.

The dynamics of the proposed network emerges during the following procedure: primarily, a graph consisting of $N$ nodes has a fully connected transition probability matrix, i.e., the probability of transition between each distinct pair of nodes is equal to $\frac{1}{N-1}$ ($\forall i,j \in \{ 1 \ldots N\},$ if $ i \neq j$ then $p_{ij}=1/(N-1) $). At every time-step, a walker selects an edge for transition; the probability of transition through the chosen edge encounters a small change. If a constructive walk happens to transit from node $i$ to node $j$, the transition probability of the selected link evolves as the following rule:

\begin{equation}\label{RuleC}
{\rm C}:\hspace{1cm} p_{ij}(t+{\delta}t)=\frac{e.p_{ij}(t)}{\sum_{m{\in}F_j} (p_{i,m}(t))+e.p_{i,j}(t)},
\end{equation}

and for destructive walkers, 

\begin{equation}\label{RuleD}
{\rm D}:\hspace{1cm} p_{ij}(t+{\delta}t)=\frac{1/e.p_{ij}(t)}{\sum_{m{\in}F_j} (p_{i,m}(t))+1/e.p_{ij}(t)}.
\end {equation}

Where equation \ref{RuleC} and \ref{RuleD} describe the dynamics of element $ij$ of the transition probability matrix as a CW or DW transits through link $ij$, respectively. $e$ is called the attachment factor; a constant which its value is essentially close to $1 ( e=1+{\epsilon}, {\epsilon}\ll1)$. $p_{ij}(t)$ is the probability of transition from node $i$ to node $j$ at time $t$. $F_{j}$ is the set of all nodes $\{m \neq j\}$, that are not occupied by any walkers. It should be mentioned that due to the normalization condition, the transition probabilities assigned to the links attached to each node must sum up to unity. Hence, the transition probabilities of the remaining links must be divided by the denominator of equations \ref{RuleC} and \ref{RuleD}, for transition of a CW and a DW, respectively. Therefore, the expectation value of transition probability, assigned to link $ij$ becomes:

\begin{equation}\label{ExpC}
p_{ij|C}=p_{ij}(\frac{(e-1)p_{ij}}{1+(e-1)p_{ij}}+\sum_{k=1}^N {\frac{p_{ik}}{1+(e-1)p_{ik}}}),
\end{equation}

and

\begin{equation}\label{ExpD}
p_{ij|D}=p_{ij}(\frac{(1/e-1)p_{ij}}{1+(1/e-1)p_{ij}}+\sum_{k=1}^N {\frac{p_{ik}}{1+(1/e-1)p_{ik}}}).
\end{equation}

Where $p_{ij|C}$ $(p_{ij|D})$ is the conditional expectation value of the transition probability for link $ij$, considering the condition that a CW (DW) has transited from node $i$. It is assumed that only one walker can pass any link at time, which is a good approximation for low walker concentration.

In the proceeding sections after an analytical evaluation, order, disorder and graph states are discussed. Afterwards, the statistics of the network is realized by examining two different weight matrices. Transition rate matrix $({\Gamma})$ is introduced as a dynamic weight matrix, describing attachment probabilities among nodes. Each element of ${\Gamma}(t)$, ${\gamma}_{ij} (t)$, is equal to the transition probability $(p_{ij}(t))$ divided by the unit time-step $({\delta}t)$, ${\gamma}_{ij}=p_{ij}/{\delta}t$. In order to investigate the traffic rate of walkers corresponding to each particular link, the traffic rate matrix ${\Omega}$ is introduced. Elements of ${\Omega}$, $({\omega}_{ij})$ are the average population transited through link $ij$, in the steady state over a sufficiently large period. Traffic rate is the measure of popularity among edges.
%%%%%%%%%%%%%%%%%%%%%%%%%%%%%%%%%%%%%%%%%%%%%%%%%%%%%%%%%%%%%%%%%
\subsection{Analytical evaluation}\label {Analytical}
The dynamics of the transition probability matrix $P(N \times N)$ can be evaluated by considering the expectation value of its elements $p_{ij}(t)$, after each transition. Considering $P_{C,i}(t)$ as the probability that a constructive walker is in node $i$, at time $t$, and $P_{D,i}(t)$ as the corresponding probability for destructive walkers, the expectation value of the evolved transition probability, $p_{ij}(t+{\delta}t)$, can be written as:
\begin{eqnarray}\label{TaylorP}
p_{ij}(t+\delta t)&=&
p_{ij|C}(t)P_{C,i}(t)+p_{ij|D}(t)P_{D,i}(t) \\ \nonumber&+&(1-(P_{C,i}(t)+P_{D,i}(t)))p_{ij}(t).
\end{eqnarray}
Here $p_{ij|C}(t)$ and $p_{ij|D}(t)$ are the conditional expectation values of transition probability, for $t+{\delta}t$ (equations \ref{ExpC} and \ref{ExpD}).
Using continuum theory~\cite{Barab99-2} for thermodynamic limit, the dynamics can be written as two coupled differential partial equations. Assuming low concentration walker approximation (at $N\rightarrow\infty$), along with the condition that attachment factor be essentially close to unity ($e=1+{\epsilon}, {\epsilon}\ll 1$), these equations can be written as (see 
\ref{AppA}):

\begin{equation}\label{PartialP}
\frac{\partial p_{ij}}{\partial t}=E(t)p_{ij}{\eta}_i(p_{ij}-M_{ii}),
\end{equation}

and

\begin{equation}\label{PartialEta}
\frac{\partial {\eta}}{\partial t}=\frac{1}{\delta t}(P^T-I){\eta}.
\end{equation}

Where ${\delta} t$ is the time interval between each transition, ${\eta}(N \times 1)$ is a vector with its elements ${\eta}_i=P_{C,i}-P_{D,i}$, and $E(t)=\lim_{\delta t, \epsilon \rightarrow 0}\frac{{\epsilon}}{\delta t}=\frac{de}{dt}(e=1)$ is a control parameter, which is assumed to be constant. $M_{ii}$ is the $i$th diagonal term of the matrix $M=PP^T$. $P^T$ is the transpose of the transition probability matrix. $I(N \times N)$ is the identity matrix. 

Equation \ref{PartialP} is a non-linear partial differential equation, which is not simply solvable analytically. According to the results which are demonstrated in the proceeding parts (see \ref{Entropy}), it is observed that when the population of CWs is less than the population of DWs, the network is given by high degree of order. This regime is called the ordered regime. As the number of DWs exceeds the population of CWs, the network happens to face a phase transition. The resulting state of the network in the steady state has relatively high disorder in transitions. This regime is called the disordered state.
A stationary state is called global, if the conditions for the links being in stationary state are satisfied for all the local components of the network. Depending on the state of the network, two global stationary states for equations \ref{PartialP} and \ref{PartialEta} are found to be attractive (see \ref{AppB}):

\emph{a.} For every node $i$, there exist a node $ j\neq i$ such that $p_{ij}=1$, and for all $k\neq j, p_{ik}=0$. If the distribution of elements of $\eta$ be uniform, this global stationary state is attractive at the ordered regime, and repulsive in the disordered state of the network.

\emph{b.} For every node $i$ and every node $j \neq i$, $p_{ij}=1/(N-1)$. If the distribution of elements of $\eta$ be uniform, this global stationary state is attractive at the disordered state, and is repulsive in the ordered state of the network.

Network reaches steady state, if $\partial {\eta} / \partial t =0$. Since every column of matrix $P$ sums up to unity, a steady state for the network is when there exist a constant ${\eta}_0<1$, such that $\forall i,j \in \{1 \ldots N\}, {\eta}_i={\eta}_j={\eta}_0$ (equation \ref{PartialEta}). In the thermodynamic limit, when every ensemble is given by same probability, if the walkers are distributed uniformly among nodes, the steady state value of $\eta$ is equal to $\eta_0=2x-1$; $x$ which is called the concentration of constructive walkers, is the ratio of constructive walkers to the total number of walkers. As $x$ is increased from $x\leq 1/2$ to $x\geq 1/2$, in the steady state, the right-hand side of equation 6 changes its sign; hence, there is a phase transition at the critical point $x=1/2$.
%%%%%%%%%%%%%%%%%%%%% Fig.1 %%%%%%%%%%%%%%%%%%%%%%%
\begin{figure*}[tp]
\vskip5mm
\includegraphics[width=15cm]{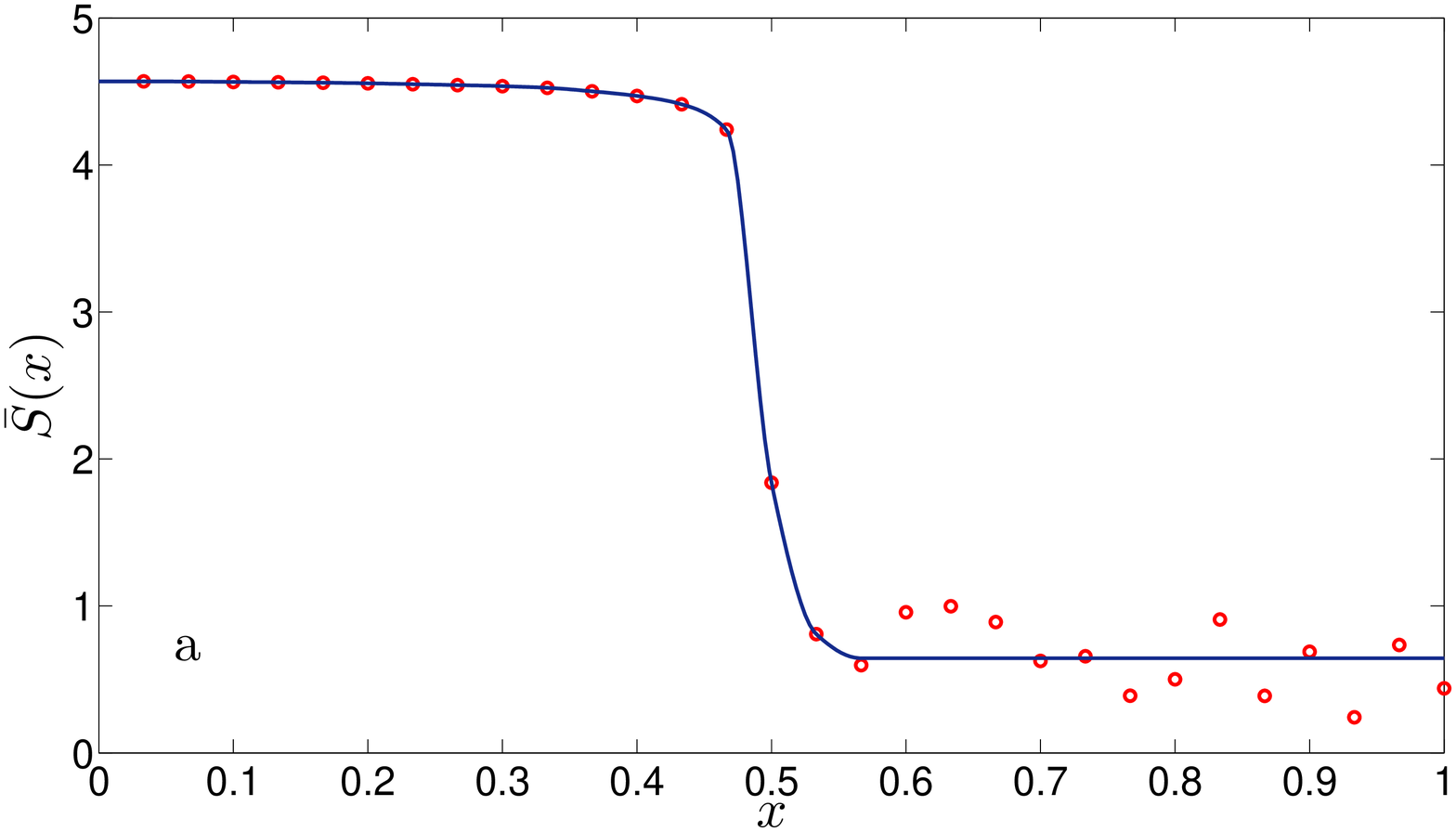}
\hspace{7mm}
\includegraphics[width=15cm]{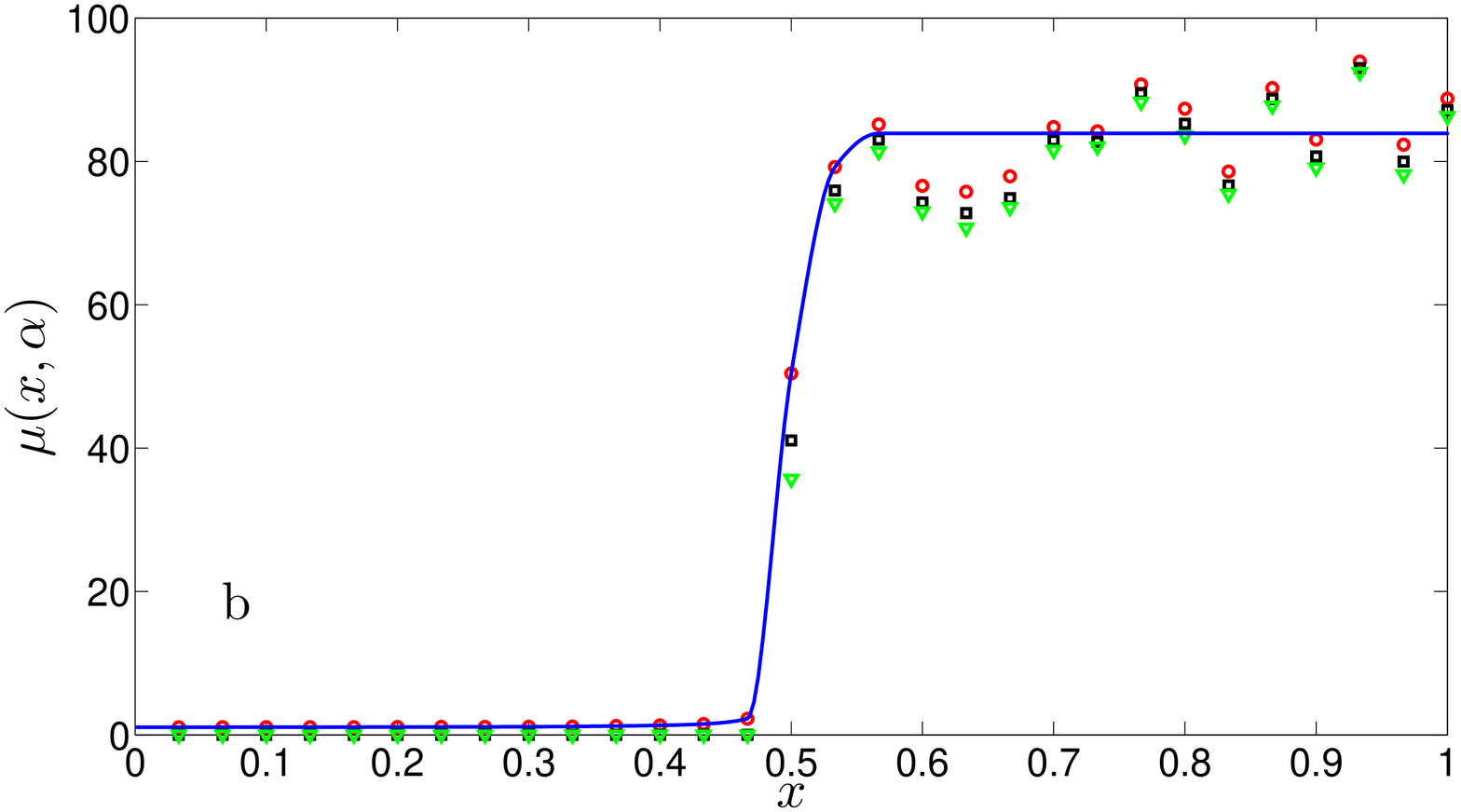}
\caption{$a)$ The average Shannon entropy as a function of constructive walkers concentration, $x$. There is a clear phase transition at $x=0.5$. b) The order parameter, ${\mu}(x,{\alpha})$, as a function of $x$ is sketched for three given values of ${\alpha}$, (red) circles for ${\alpha}=2$, (black) squares for ${\alpha}=4$, and (green) triangles for ${\alpha}=8$. The graph contains $N=100$ nodes and $m=30$ walkers, with $e=1.1$. The results are obtained after the systems approached steady state. The solid curves are plotted just as eye guides.}
\label{fig1}
\end{figure*}
%%%%%%%%%%%%%%%%%%%%%%%%%%%%%%%%%%%%%%%%%%%%%%%%%%%%%%%
%%%%%%%%%%%%%%%%%%%%% Fig.2 %%%%%%%%%%%%%%%%%%%%%%%
\begin{figure}[tp]
\vskip3mm
\includegraphics[width=15cm]{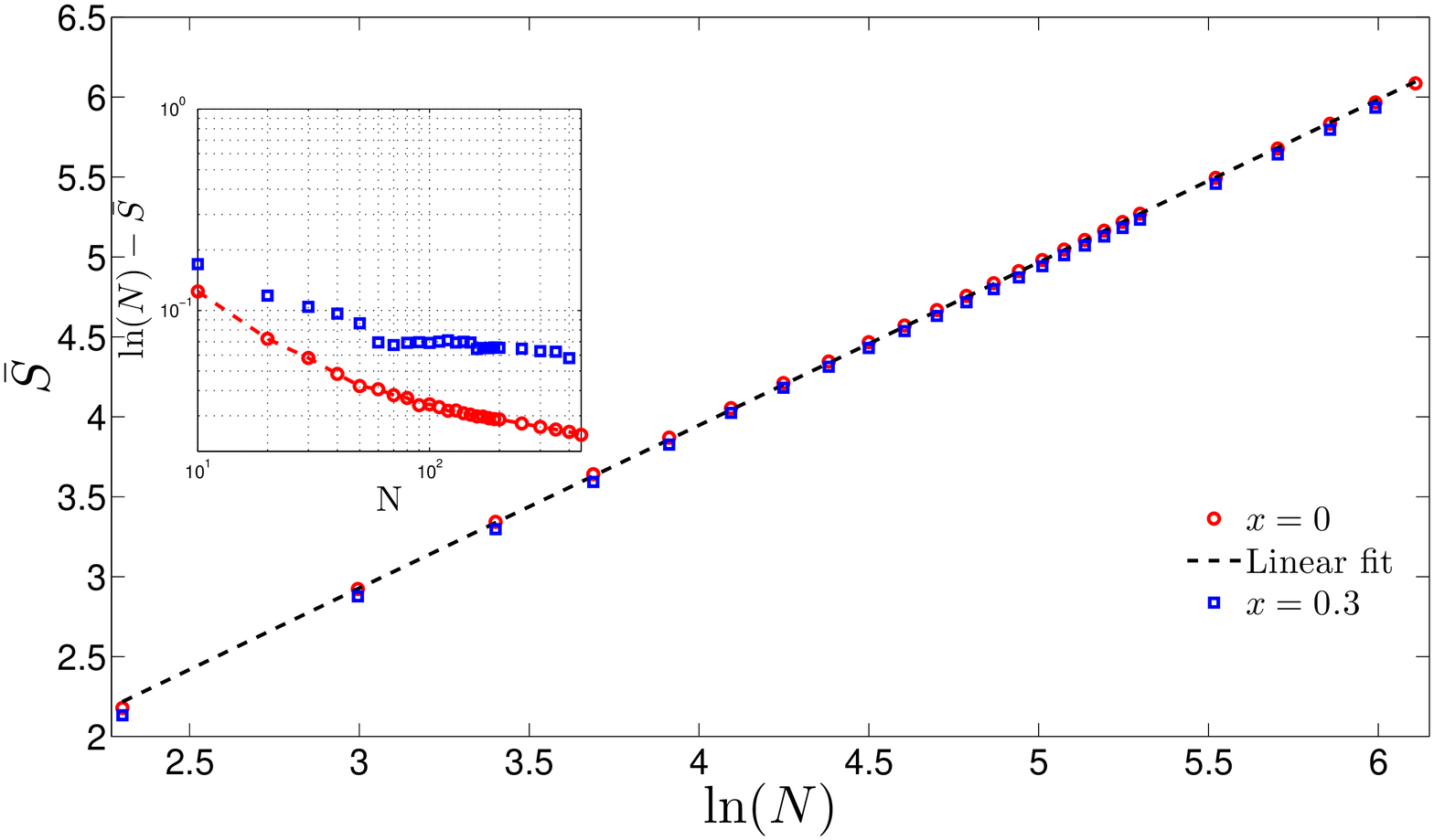}
\caption{Average Entropy for fixed walker concentrations shows logarithmic growth with the network size. Here, the results for two concentrations, $x=0$ (red) circles, and $x=0.3$ (blue) squares, are presented. The inset diagram shows deviations from the logarithm of network's size.}
\label{fig2}
\end{figure}
%%%%%%%%%%%%%%%%%%%%%%%%%%%%%%%%%%%%%%%%%%%%%%%%%%%%%%%%%%%%%%%%%
%%%%%%%%%%%%%%%%%%%%% Fig.3 %%%%%%%%%%%%%%%%%%%%%%%
\begin{figure}[bp]
\vskip4mm
\includegraphics[width=15cm]{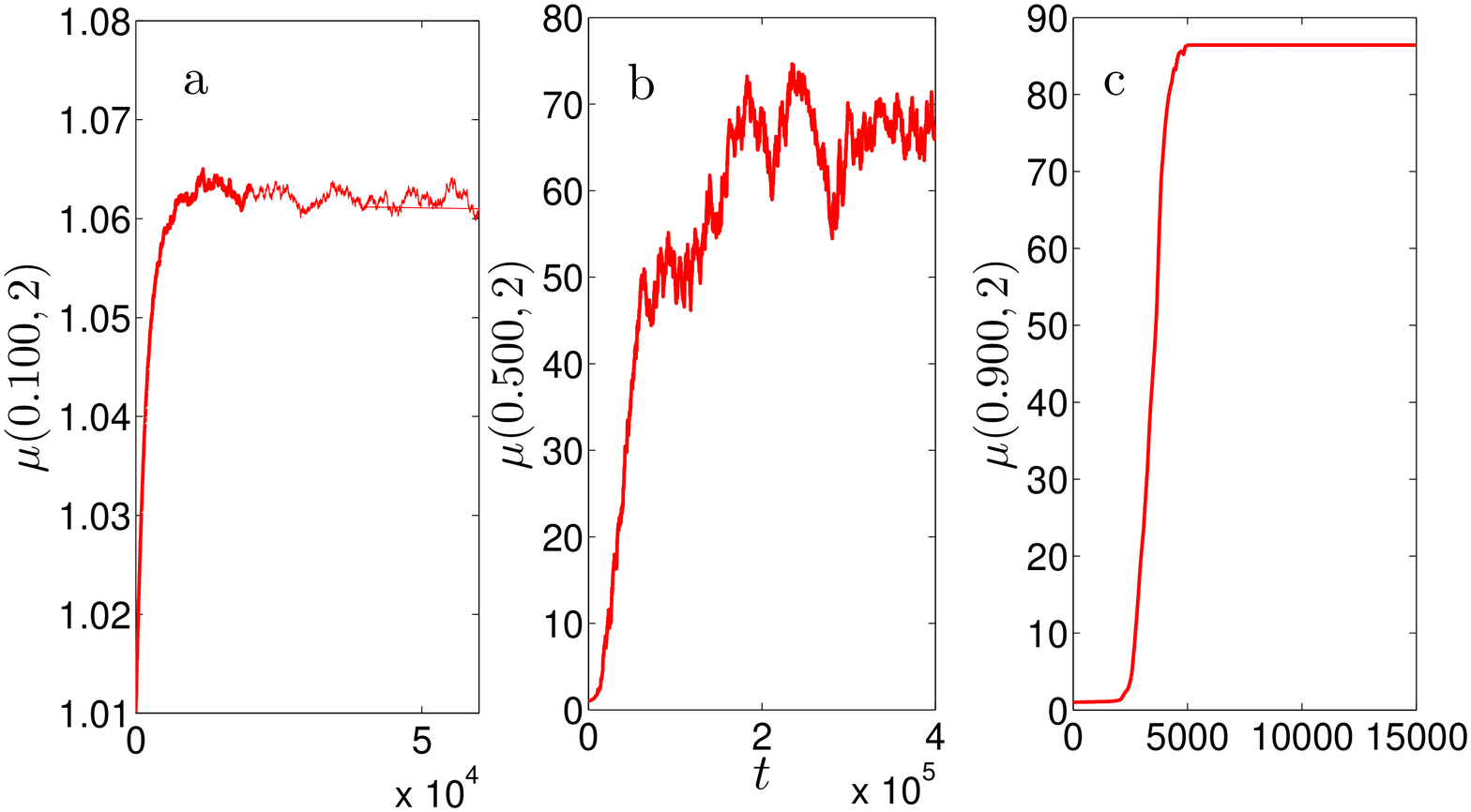}
\caption{Order parameter for $\alpha=2$ in three observed states of the network; a) the disordered state with $x=1/30$, b) the transition state with $x=15/30$, and c) the ordered state of the network, with $x=27/30$. The system shows a slow down relaxation in transition state (as expected). The relaxation time for (c) is $t \approxeq 5000$ times-steps and for (a) is $t \approxeq 10000$. But it is not relaxed in transition state, even after $t=400000$ time steps.}
\label{fig3}
\end{figure}
%%%%%%%%%%%%%%%%%%%%%%%%%%%%%%%%%%%%%%%%%%%%%%%%%%%%%%%
\subsection{ Entropy, order parameters and phase transition}\label{Entropy}
The average Shannon entropy~\cite{Jean2005} quantifies the average uncertainty a walker has in selecting an edge, as it happens to transit from a randomly selected node. This quantity which determines the correlation between disorder in network and the types of walkers is written as:
\begin{equation}
\bar{S}(x)=\frac{1}{N}\sum_{i,j}p_{ij}ln(p_{ij}).
\end{equation}
$p_{ij}$ is the probability of transition from node $ i$ to node $j$, in the steady state. As stated previously, the argument $x$ (concentration of constructive walkers) is the ratio of CWs' population ($n_c$) to the total number of walkers $(m)$, $x=\frac{n_c}{m}$ .

As the transition probabilities for edges connected to one node sum up to unity $\sum_{j}P_{ij}=1$, then $\sum_{ij}P_{ij}=N$. If we introduce the order parameter as the following:
\begin{equation}\label{OrderParameter}
{\mu}(\emph{x},{\alpha})=\sum_{ij}p_{ij}^{\alpha}.
\end{equation}

For ${\alpha}>1$ it counts the influence of high weighted links. For ${\alpha}\rightarrow\infty$, ${\mu}$ approaches to the number of links that has a probability equal to 1.

The model for a graph containing $N=100$ nodes, and a fixed total number of walkers $m=30$, with factor $e=1.1$, is simulated for several concentrations of CWs. The simulation results for the above-mentioned parameters reveal that in agreement with results from analytical evaluation (\ref{Analytical}), there is a clear condensation~\cite{Dor} phase transition, when the populations of the two types of walkers are equal, $x=0.5$. The resulting plots corresponding to the average entropy and several order parameters $({\alpha}=2, 4, 8)$ are depicted in Figure \ref{fig1}. Depending on the value of $x$, the graph is categorized in three states: the ordered state (O-State: $x>0.5$), the disordered state (D-State: $x<0.5$), and the transition state (T-State: $x\approxeq 0.5$). As the curve shows, the system goes ordered with low entropy, at $x>0.5$; and transits to the D-State for $x<0.5$, which is given by relatively high entropy. This shows, however, while constructive walkers try to build a few high-traffic links in the networks, the destructive walkers tend to make them uniform.

In order to examine the functionality of the average entropy with respect to the size of the network, this quantity is calculated for several network sizes, for several concentration of CWs. The simulation data reveals that the average entropy of the networks with small $x$ grows logarithmically with respect to the network's size. This result is depicted in Figure \ref{fig2}. The inset diagram of Figure \ref{fig2} is the deviation of average Shannon entropy of the network from natural logarithm of network's size ($\ln(N)-\bar{S}$). According to this result average Shannon entropy has a dominant term equal to $\ln(N)$, and as $x$ is increased, deviation from this dominant term increases.
%%%%%%%%%%%%%%%%%%%%% Fig.5 %%%%%%%%%%%%%%%%%%%%%%%
\begin{figure*}[tp]
\vskip5mm
\includegraphics[width=15cm]{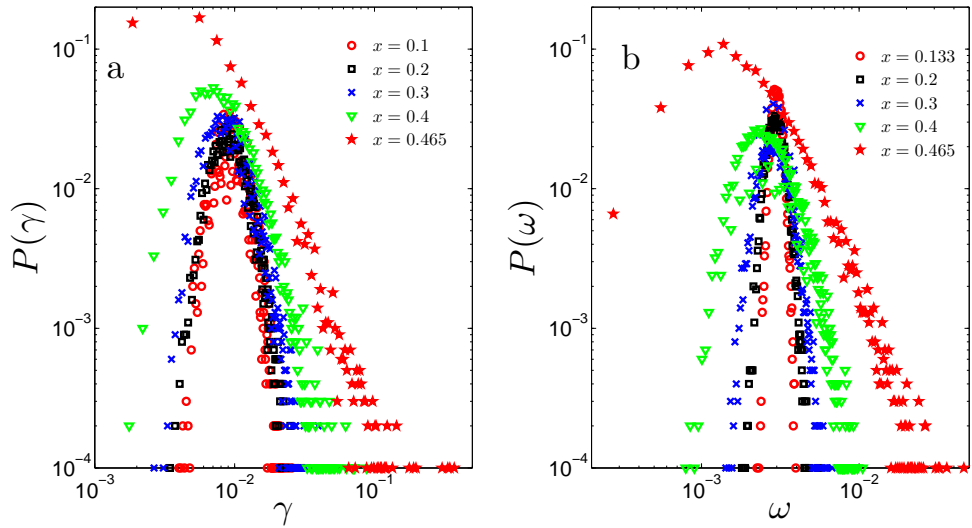}
\caption{Transition rate (a) and traffic rate (b) distributions for some different values of concentrations, $x \in [0,0.5]$. It is clear that as $x$ approaches to its critical point $x=0.5$, scaling in the weight distributions emerges.
}
\label{fig6}
\end{figure*}
%%%%%%%%%%%%%%%%%%%%%%%%%%%%%%%%%%%%%%%%%%%%%%%%%%%%%%%
%%%%%%%%%%%%%%%%%%%%% Fig.4 %%%%%%%%%%%%%%%%%%%%%%%
\begin{figure}[tp]
\vskip4mm
\includegraphics[width=15cm]{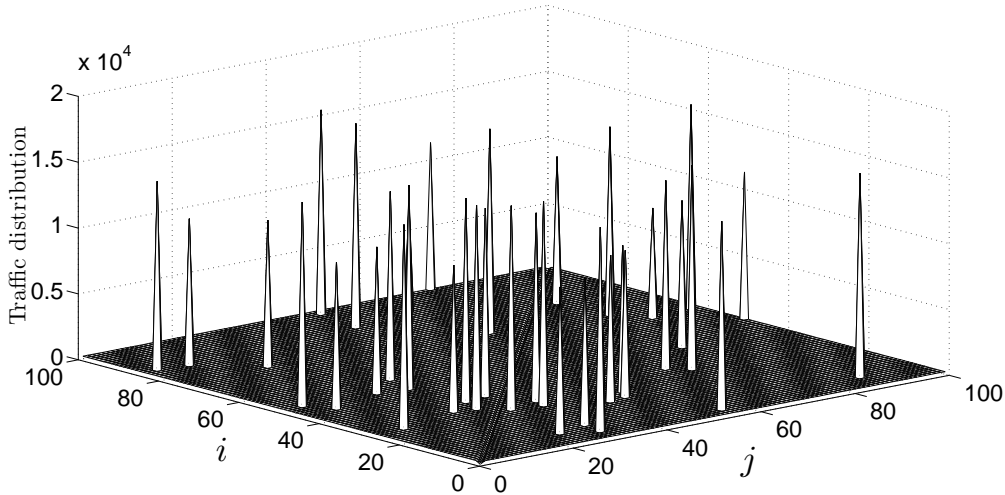}
\caption{The traffic distribution on the network, in the steady state, for the ordered state in a $100\times100$ network. The vertical axis is the population transited from the node $i$ to node $j$.}
\label{fig5}
\end{figure}
%%%%%%%%%%%%%%%%%%%%%%%%%%%%%%%%%%%%%%%%%%%%%%%%%%%%%%%
%%%%%%%%%%%%%%%%%%%%% Fig.6 %%%%%%%%%%%%%%%%%%%%%%%
\begin{figure}[tp]
%\vskip5mm
\includegraphics[width=14cm]{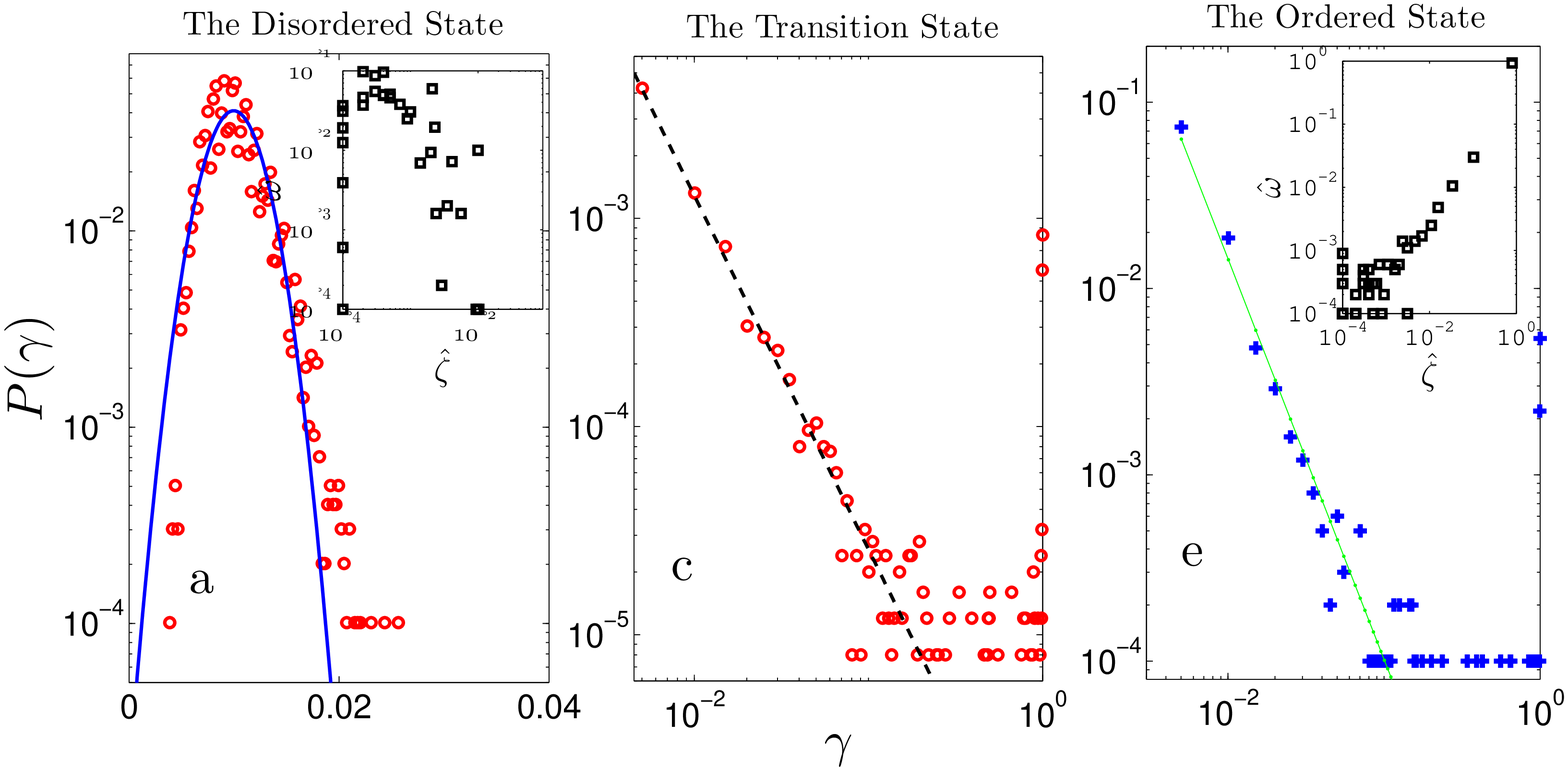}
%\hspace{7mm}
\includegraphics[width=14cm]{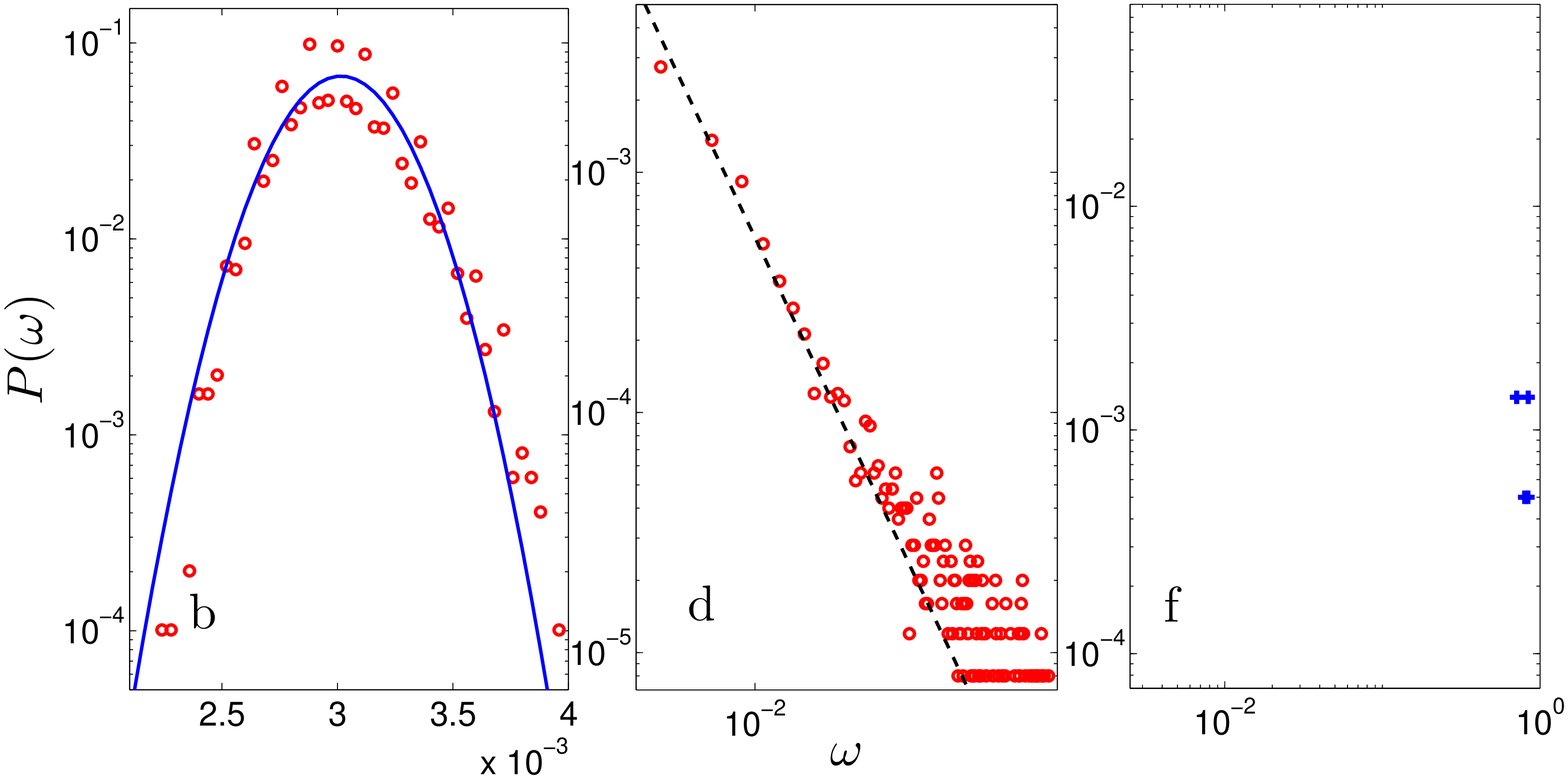}
\caption{Weight distributions for the three states of the network are sketched. The total number of walkers to the size of the network are kept fixed on $m/N= 0.3$. In (a) and (b) $N=100$ and $x=3/30$, in (c) and (d) $N=500$ and $x=75/150$, and in (e) and (f) $N=100$ and $x=27/30$. In ordered state, while we see a scale free behavior for transition rates, the traffic is localized in a few alive links. On the other hand, both the rates show normal distributions in the disordered phase. Power-law distributions of the rates in the transition state show that the network in its critical point, is scale free. The best power-law fits on the data are shown with black lines with the slope of $-1.7$ for (c) and $-1.4$ for (d). The linear branch of transition distribution for ordered state also fits best on a power-law with a slope of $-2.15$. The inset shows the comparison diagram for the rates. As one can see, there are a strong correlation between the rates in transition state with a regression equal to $R=0.994$.}
\label{fig4}
\end{figure}
%%%%%%%%%%%%%%%%%%%%%%%%%%%%%%%%%%%%%%%%%%%%%%%%%%%%%%%
%%%%%%%%%%%%%%%%%%%%% Fig.9 %%%%%%%%%%%%%%%%%%%%%%%
\begin{figure*}[tp]
\vskip5mm
\includegraphics[width=15cm]{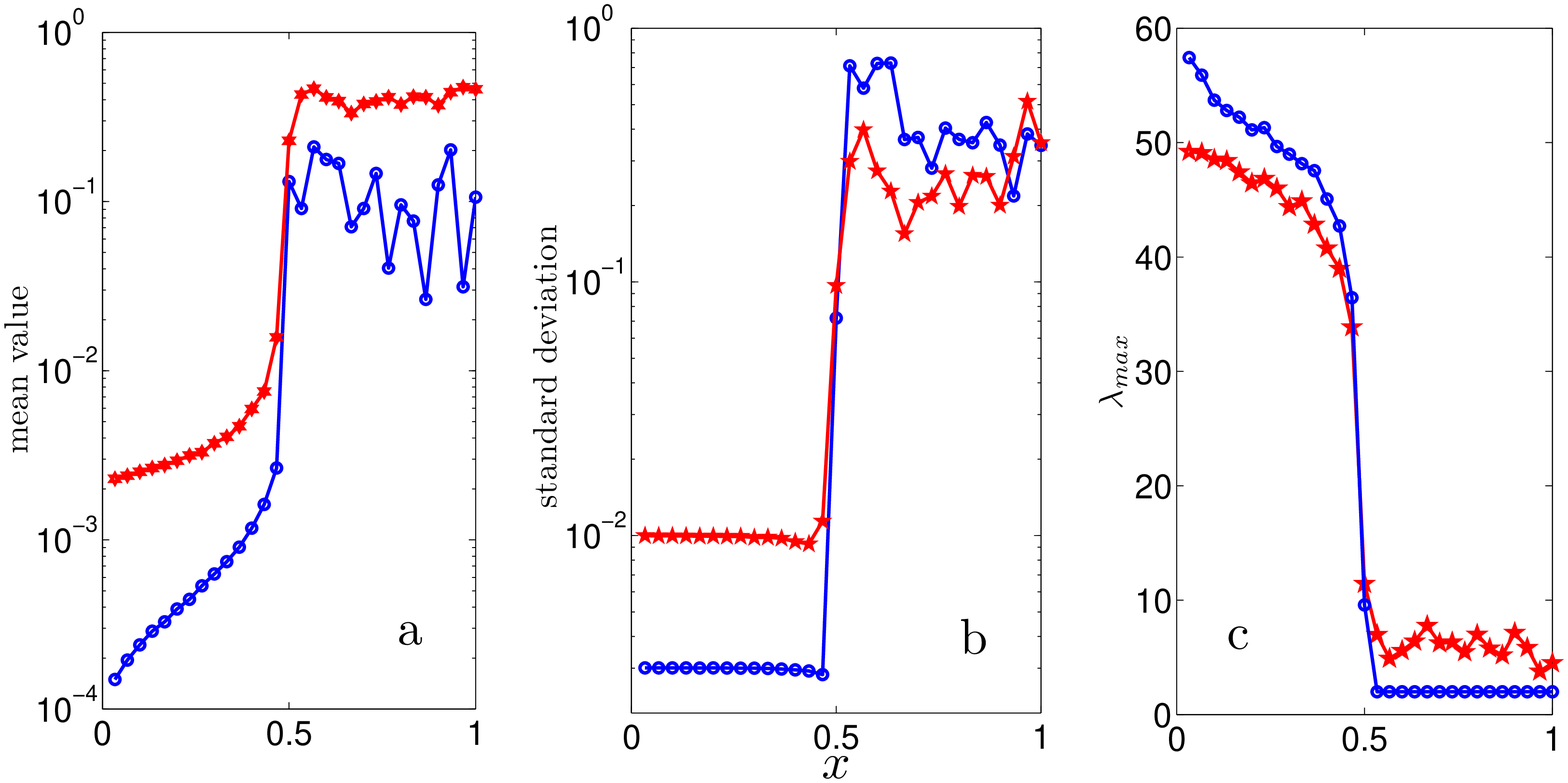}
\caption{Statistical properties of the network, as a function of concentration of constructive walkers. In all the graphs, (red) stars show the result for transition rate, and (blue) circles are for traffic rate. In (a) the average rates are shown as a function of $x$, (b) shows the standard deviation for the rate distributions, and (c) shows the principal eigenvalue of the adjacency matrix.
}
\label{fig9}
\end{figure*}
%%%%%%%%%%%%%%%%%%%%%%%%%%%%%%%%%%%%%%%%%%%%%%%%%%%%%%%
\section{The statistical properties}
The statistics of transition rate and traffic rate among links, in the three previously stated regimes, are evaluated using the weight distributions, spectral densities, average clustering coefficient, and average path length. We used several transition rate and traffic rate distributions to demonstrate the directional connectedness for each edge and popularity among edges, respectively ~\cite{Barth2004}. To extract the results, we wait until the network approaches to the steady state. We suppose the system in the steady state, when the time-dependent second-order parameters ($\mu(x,2)$), as macroscopic variables indicating the influence of highly weighted links, are saturated. As it is demonstrated in Figure \ref{fig3}, the at the critical point of transition (T-State) system does not relax even after a long period of time. For the O-State and D-State the relaxation in network is after a finite time steps. According the ratio of the different walkers in the network, the system converges to three different states; the ordered, the disordered, and the transition states:

\textbf{D-State ($x>0.5$)}. This regime is constructed by dominant effect of DWs. The properties of the network for this state are mostly like an almost connected network, with distributions alike to a random graph. The fully connected network in this regime is stable.

\textbf{T-State ($x\approxeq 0.5$)}. In this regime, the number of both types of walkers is the same $(x=0.5)$. As the CWs transit through $m/2$ of edges, these links will be given by increased transition probabilities, and therefore destructive walkers, which have equal populations $n_D=n_C$, will be absorbed to these edges. Hence, the walkers interact so strongly. Figure \ref{fig6} demonstrates the emergence of scaling, as $x$ increases from small values ($x<0.5$) to $x=0.5$.

\textbf{O-State ($x>0.5$)}. This state of the network is constructed by dominant effect of CWs. Most of the traffic transferred within the network, is distributed among a few links. Since $x>0.5$, a positive value for ${\eta}_i$ is expected, and the initial fully connected network is repulsive. The CWs in the system tend to direct the traffic through a few graph links and therefore, destroy the rest, and as a consequence, detached sub-graphs appear. As demonstrated in Figure \ref{fig5}, in the steady state, the traffic is located among a few numbers of edges.

\subsection{ Weight distributions} Weight distributions for the three states of the network are demonstrated in Figure \ref{fig4}. For both traffic and transition rates in the D-State, the distributions are alike to a \emph{Gaussian} distribution. As $x$ is increased from $x=0$ to $x=0.5$, a power-law tail for the distributions appears (Figure \ref{fig6}). The destructive walkers tend to construct a random graph. As the network transits from D-State to the T-state ($x=0.5$), the distributions for both transition rate and traffic rate take a power-law tail with factors 1.7 and 1.4, respectively. Transition rate distribution at this critical point ($x=0.5$) includes several singular spots for high weights in the tail. The transition rate distribution in the O-State, includes a linear branch with power-law distribution with an exponent varying between 1.7 and 2.2. As the value of CWs concentration is increased, the slope of the power-law branch increases monotonically. At the end of the tail of the power-law there are many singularity spots related to highly weighted links. The traffic rate consists distribution consists of highly weighted singular links.

In order to investigate the similarity among transition rate and traffic rate distributions, the linear dependency between them is examined. Considering weight matrix $W$ assigned to the network, the number of links with their weights $w=w_0$ is written as $N(w=w_0)$. A normalized weight is defined as $\hat{W}=W/\max{(W)}$ . The inset diagrams of Figure \ref{fig4} are called the comparison diagrams. Each point in the insets corresponds to a particular value of $\hat{\gamma}=\hat{\omega}$. The vertical axis is the number $(N(\hat{\omega}))$ of links associated to normalized traffic rate $(\hat{\omega})$, and the horizontal axis is the number $(N(\hat{\gamma}))$ of links associated to the normalized transition rate $(\hat{\gamma}=\hat{\omega})$. Traffic rate distribution in the O-State consists of several singularities, associated to highly weighted links. The transition rate distribution, along with the singular links, includes a linear branch in logarithmic scale. The calculated regression for the comparison diagrams in the D-State and the T-State are $R=0.298$ and $R=0.994$, respectively. This result shows that for the T-State of the graph, both distributions take similar shapes; for D-State, although both distributions are fitted to a Gaussian, their curve in comparison diagram deviates from line.
%%%%%%%%%%%%%%%%%%%%%%%%%%%%%%%%%%%%%%%%%%%%%%%%%%%%%%%%%%%%%%%%%
\subsubsection{Constructive walkers and Network Branching} 
\begin{figure}[bp]
\vskip4mm
\includegraphics[width=15cm]{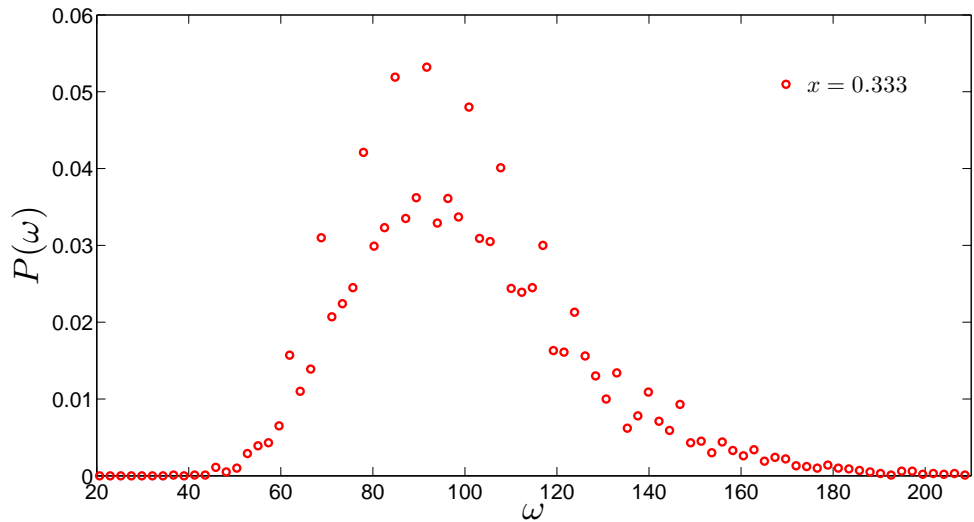}
\caption{ The traffic rate distribution is sketched for disordered state at $x=3/30$ for a network consisting of $N=100$ nodes. The distribution consists of two clear branches. This branching is due to the correlations among constructive walkers and destructive walkers (see the text).}
\label{fig7}
\end{figure}

By the dominant effect of DWs, the network tends to become fully connected. As $x$ is increased, the variance for distributions expands monotonically (see Figure \ref{fig9}), and for a pure DW population ($x=0$) the distribution's shape is a simple Gaussian. According to simulation results from the D-State, as the population of CWs increases, the traffic rate distribution deviates from a simple Gaussian. As it is demonstrated in Figure \ref{fig7}, the distribution branches into two overlapping Gaussian distributions, with same averages and standard deviations satisfying $\frac{(|{\sigma}_2-{\sigma}_1 |)}{(({\sigma}_1+{\sigma}_2)/2)}\ll1$; with ${\sigma}_i$ ($i=1,2$) being the standard deviations for distribution branches. We call this behavior the network branching. This phenomenon is due to the correlations between the CWs and the DWs. As the population of CWs increases, interactions between walkers become significant. Hence, the CWs interact with a couple of DWs. When CWs pass certain links, a number of DWs are absorbed to these edges. As the DWs transit through these links, the transition probability for the selected links reduces, and the rest of the DWs transit through fewer selective edges. Therefore, the network's geometry crashes into two weak interacting sub-graphs.
%%%%%%%%%%%%%%%%%%%%%%%%%%%%%%%%%%%%%%%%%%%%%%%%%%%%%%%%%%%%%%%%%
\subsection{ The spectral properties}
To every network $(G)$, we can assign an adjacency matrix $A(G)$, which represents the connectivity between vertices. Adjacency matrix is a ${N}\times{N}$ matrix, with elements $A_{ij}=A_{ji}$; nodes $i$ and $j$ are connected if and only if $A_{ij}=1$ and $A_{ji}=1$. For a network consisting of $N$ nodes, the adjacency matrix has $N$ eigenvalues ${\lambda}_{j},j\in\{1\ldots{N}\}$. The spectral density is the distribution of the eigenvalues. In mathematical terms, the spectral density is defined as:
\begin{equation}
{\rho}({\lambda})=\frac{1}{N}\sum_{i=1}^{N}{\delta}({\lambda}-{\lambda}(i)).
\end{equation}
This summation as $N\rightarrow \infty$, approaches to a continuous function. As Wigner ( ~\cite{Wigner55, Wigner57, Wigner58}) had stated, for the case of a random graph of $N$ nodes with $p$ as the probability of attachment, if  $|{\lambda}|<2\sqrt{Np(1-p)}$ the eigenvalue distribution takes a semi-circular shape as:

\begin{equation}
\rho(\lambda)=\frac{\sqrt{4Np(1-p)-{\lambda}^2}}{2{\pi}Np(1-p)}.
\end{equation}

The value of $\rho$ for $|{\lambda}| \geq 2\sqrt{Np(1-p)}$ is zero, except for several eigenvalues localized far from the center of the semi-circle. This is also known as the semi-circular law. The largest eigenvalue is called the principal eigenvalue, which plays an important role in topological features of the network~\cite{Barab2002}.

According to several works ~\cite{Farkas2001, Kim2001} for scale-free networks, the spectral density takes a triangular shape, with its top lying high above the semicircle. Due to the nodes localized with high degrees in a scale-free network, the margins decay as a power-law. In scale-free networks, the principal eigenvalue defines the network's loop structures.
%The logic behind this definition is as the following: with respect to the defined weight, node $i$ is attached to node $j$ %with a strength proportional to $W_{ij}$, and node $j$ is connected to node $i$ with a strength proportional to %$W_{ji}$. On average, the non-directional strengths for both vertices are attached together with strength proportional %to $(W_{ij}+W_{ji})/2$.
%%%%%%%%%%%%%%%%%%%%% Fig.7 %%%%%%%%%%%%%%%%%%%%%%%
%%%%%%%%%%%%%%%%%%%%%%%%%%%%%%%%%%%%%%%%%%%%%%%%%%%%%%%
%%%%%%%%%%%%%%%%%%%%% Fig.8 %%%%%%%%%%%%%%%%%%%%%%%
\begin{figure*}[tp]
\vskip5mm
\includegraphics[width=15cm]{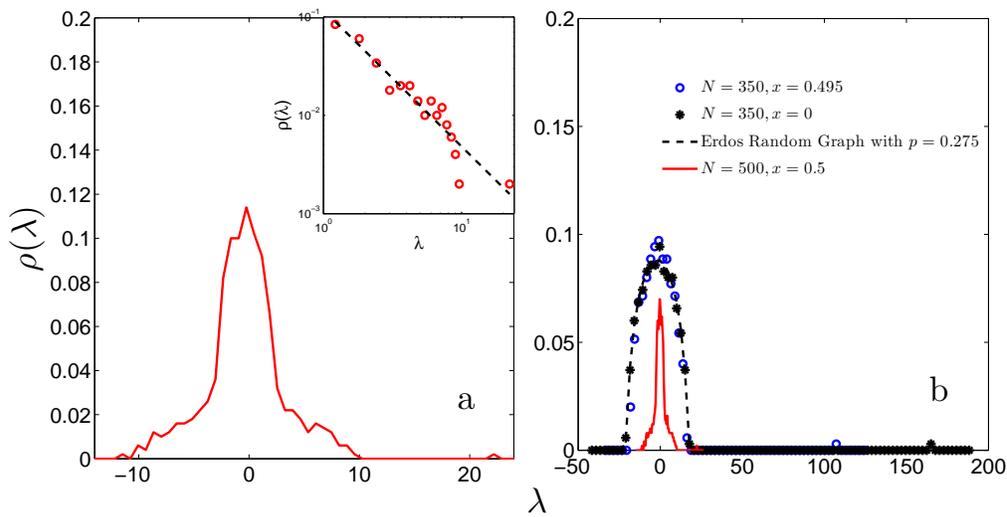}
\caption{The spectral density transition rate (a) for a network in its critical point ($x=0.5$). The network consists of $N=500$ nodes with $m=150$ random walkers. The inset depicts the margin in log-log. It is compared with the spectral densities for the network in ordered and disordered states in (b). Also a solid curve compares it with the best fit to the spectrum of an Erd{\H o}s graph. }
\label{fig8}
\end{figure*}
%%%%%%%%%%%%%%%%%%%%%%%%%%%%%%%%%%%%%%%%%%%%%%%%%%%%%%%
We used a rounding procedure to assign an adjacency matrix to the network. For $W$ as a weight matrix, the symmetric part of $W$ is defined as $W^s=\frac{1}{2}(W+W^T)$, which $W^T$ is the transpose of $W$. This symmetric weight, represents the average non-directional strengths of connectedness among nodes. The adjacency matrix $A(W)$ is constructed by comparing $W^s$ with average symmetric weight ${\bar{W}}^s=(2/N(N-1) )\sum_{i,j}W^s_{i,j}$ , i.e., if $W^s_{ij}\geq\bar{W^s}$ the corresponding adjacency matrix is assigned to be 1 $(A(W)_{ij}=A(W)_{ji}=1)$, and otherwise 0.
%%%%%%%%%%%%%%%%%%%%%%%%%%%%%%%%%%%%%%%%%%%%%%%%%%%%%%%%
\subsubsection{Results}
We have used the symmetric traffic rate and transition rate matrices. For the T-State, the eigenvalue distribution is a triangular shape, and the topology of the network is scale-free (see Figure \ref{fig8}-a). The results for the D-State (Figure \ref{fig8}-b), reveal that the spectral density is a semi-circular shape, and the topology of the network is similar to a random graph. By comparing the results for the network in the D-State with an Erd{\H o}s network, it is observed that the spectral density function fits to a random graph with attachment probability equal to $p=0.275$. In order to demonstrate the topology transition as the concentration $x$ evolves, the behavior of the principal eigenvalue, average weights and standard deviation of weights from the mean amount are depicted as a function of $x$ in Figure \ref{fig9}.
%%%%%%%%%%%%%%%%%%%%%%%%%%%%%%%%%%%%%%%%%%%%%%%%%%%%%%%%%%%%%%%%%

\subsection{ Clustering and average path length}Further topological features of the network, in the three states, are examined using average clustering coefficient and average minimum path. In the following, after indicating our approach, the results corresponding to average clustering coefficient and average distance between nodes in several CWs concentrations are demonstrated.

\subsubsection{Clustering} Clustering coefficient is the measure that determines how nodes in the neighborhood of a specified node interact together. In general, this measure is equal to the fraction of triangles surrounding a node to the total potential triangles within the node's neighborhood.
Recently, clustering coefficient has been modified for weighted networks. Depending on the modeling, there are several candidates for the definition of clustering coefficient. Several definitions have been introduced, that through their definitions, the neighborhood structures, adjacency matrix or strength of the nodes is used to define the clustering coefficient (Lopez-Fernandez (2004), Barrat, Onnela ~\cite{Barrat2004,Wen2005}). In some definitions, since the definition requires an underlying binary network to be generated, rounding procedures concerning threshold parameters are used (Rougemont and Hingamp 2003 ~\cite{Barth2004,Bocc2006, Newman2001-1,Kanla2000,Jari2006}). We have focused on the definition of Zhang and Horvath, which was traced back to the earlier work of Grinlord. The advantage of this procedure is that it deals directly with normalized weights, and does not need the weights to be rounded to 0 or 1. Furthermore, in the procedure, the clustering coefficient is defined only by using the weights.

In Zhang-Horvath's definition, for $W$ as the weight matrix assigned to the network, the weights are normalized to $\max{(W)}$, and the clustering coefficient for node $k$ is written as:
\begin{equation}
C_{HZ}(k)=\frac{\sum_{j,k}\hat{W}_{ij}\hat{W}_{jk}\hat{W}_{ki}}{\sum_{j\neq{k}}\hat{W}_{ij}\hat{W}_{ki}}
\end{equation}
Where $\hat{W}$ is the normalized weight matrix.
%%%%%%%%%%%%%%%%%%%%%%%%%%%%%%%%%%%%%%%%%%%%%%%%%%%
\subsubsection{Average minimum path}
By definition, the distance from node $i$ to node $j$ in the graph is the minimum number of possible connected links that originate from node $i$ and terminate to node $j$. Minimum path can be defined using normalized weights. We used the Floyd-Warshall Algorithm ~\cite{Howard} for weighted networks. The algorithm consists of three steps:
%%%%%%%%%%%%%%%%%%%%% Fig.10 %%%%%%%%%%%%%%%%%%%%%%%
\begin{figure*}[tp]
\vskip5mm
\includegraphics[width=15cm]{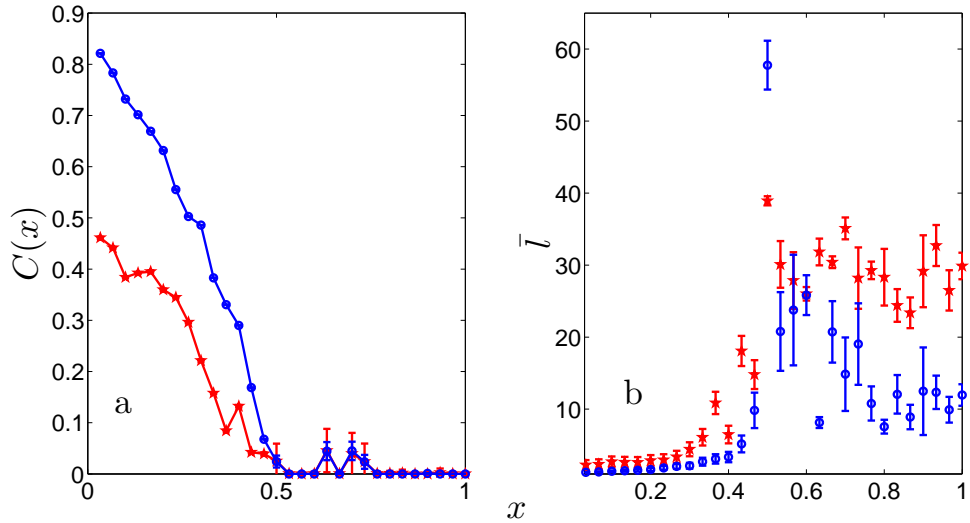}
\caption{The average clustering $C(x)$ and the average path length $\bar{l}$ as a function of $x$ for a network consisting of $N=100$ nodes with $m=30$ and $e=0.1$ (a). For both traffic rate, (blue) circles, and transition rate, (red) stars, network has high clustering feature in the disordered state and as $x$ is increased, average clustering coefficient decreases monotonically. In ordered state almost for all concentrations network has trivial average clustering. But the average path length shows is peaked for both the rates around the transition state. To find the average path length in ordered states we just looked at inside the isolated islands.}
\label{fig10}
\end{figure*}
%%%%%%%%%%%%%%%%%%%%%%%%%%%%%%%%%%%%%%%%%%%%%%%%%%%%%%%
1) For $W (N\times N)$ as the weight matrix, every weight is normalized to $\max{(W)}$, and the distance between every adjacent pair of nodes $i,j$ is defined as $d_{ij}=\frac{1}{\hat w _{ij}}$ , where $\hat{W}$ is the normalized weight matrix.
2) The diagonal elements of the distance matrix are assigned zero $(diag(d)=0)$.
3) The distance between every pair of nodes $i,j$ is assigned $min_{k\in 1 \ldots N}$ ($d_{ij}$,$d_{ik}+d_{kj}$). This step is repeated iteratively for $N$ times, and the final $d$ matrix is the minimum path matrix. 
The normalized traffic rate and transition rate matrices are used to specify the properties of the network with respect to clustering and average minimum paths.
\subsubsection{Results}
Figure \ref{fig10} depicts the results regarding average clustering and average path length of the network. In the D-State with small $x$, both the traffic rate and transition rate networks are highly clustered. As $x$ increased, the average clustering coefficient decreases monotonically. In the T-State and the O-State, the average clustering of the network is trivial. The characteristic length for both transition rate network and traffic rate networks in the D-State is small compared to the network's size. As $x$ is increased, the average path length increases. In the T-State, the average minimum path reaches to a finite large amount. The O-State consists of several isolated sub-graphs, and the characteristic length for both traffic rate and transition rate networks is infinite. For these states, we have considered the average distance for each isolated clusters. The average over the clusters' average distances is considered as the average path length of the network. The average minimum path in the O-State and the T-State, are high above the average distances in the D-State.
%%%%%%%%%%%%%%%%%%%%%%%%%%%%%%%%%%%%%%%%%%%%%%%%%%%%%%%
\section{Summary}
We proposed a model of dynamic network with interactive random walkers. We consider the walkers to be of two classes, not interacting with each other, but have a trace on the links of the network. The walkers of the first type, which cause an enhancement in links' transition probabilities, are called the constructive walkers (CW), and the others, which cause a reduction in the transition probabilities, are called the destructive walkers (DW). In the limit of low walker concentration approximation, the analytical evaluation of the dynamics of the transition probability matrix results in two coupled partial non-linear differential equations (equations \ref{PartialP} and \ref{PartialEta}).
Depending on the relative concentration of the two types walkers, $x$, there are two different global stationary states for the network. When the constructive ones are dominant, the system approaches a singular network (O-state), with a few non-vanishing links which the traffic is concentrated on them. On the other hand, when the number of destructive walkers is higher, the network tends to be uniform (D-state). In the between, when the network is given by a balanced population of the walkers, a sharp transition from singular to uniform state is observed. The system in the transition state (T-state) shows clear characteristics of scale-free networks. 
To distinguish the phase transition, we introduce an order parameter ( equation~\ref{OrderParameter}) which is tiny for an uniform network and counts the number of high traffic paths on the singular network.
Also, we are able to look at average Shannon entropy of the network which accounts randomness in transition rates. As the network size increases, the average Shannon entropy in the D-State has a dominant term equal to the natural logarithm of network's size ($S(N)\varpropto \ln(N)$), and as $x$ is increased, the average entropy deviates from this functionality. We introduced the transition rate matrix, which its elements are the transition rate of the walker through the links, indicating the connectedness among nodes. The traffic rate matrix, which its elements are the average population transited through links per unit of time-step (during a long period after the steady state), gives the popularity among edges. In D-State, both the transition and traffic rate distributions are almost Gaussian, with a spectral density well fitted to an Erd{\H o}s random graph. By increasing the concentration of CWs, the traffic distributions branches into two overlapping Gaussian distributions. We called this phenomenon, which is caused by correlations between the walkers, the network branching.
Regarding the spectral densities and the wight distributions, we found that network in the T-State is scale-free, with exponents $-1.7$ and $-1.4$ for transition and traffic rates, respectively. In the O-State also we see marginally scale-free behavior on transition rates with slope of $-2.2$. In this case, the traffic is catches in a few closed paths and is isolated from the rest of the network. Then, it does not show scale-free structure in steady state. The results regarding clustering and characteristic length of the network, reveal that the D-State has high clustering, and the short path lengths; while for the O-State and the T-State network consists of isolated clusters, and the average over the cluster's path lengths were observed to be high above the average lengths in the D-State.
The introduced model can be applicable to many realistic models, and can explain the scale free features of dynamical networks, interacting with their elements.
%%%%%%%%%%%%%%%%%%%%%%%%%%%%%%%%%%%%%%%%%%%%%%%%%%%%%%%%%%%%%%%%
\ack
We thank Reza Jafari for very useful comments, and also Amir Najafi for his valuable hint on analytical calculation. We also thank the Center of Excellence in Complex Systems and Condensed Matter (CSCM) for partial support.
%%%%%%%%%%%%%%%%%%%%%%%%%%%%%%%%%%%%%%%%%%%%%%%%%%%%%%%%%%%%%%%%
\appendix
\section{The dynamics of transition probability matrix}\label{AppA}
Using Taylor expansion, considering the first non-zero terms of $\epsilon$, equations \ref{RuleC} and \ref{RuleD} will become:
\begin{equation}
p_{ij|C}=p_{ij}(p_{ij}(1+\epsilon)-\epsilon p^2_{ij}+\sum_{k \neq j}p_{ik}-\epsilon \sum_{k\neq j} p^2_{ik}),
\end{equation}
\begin{equation}
p_{ij|D}=p_{ij}(p_{ij}(1-\epsilon)+\epsilon p^2_{ij}+\sum_{k \neq j}p_{ik}+\epsilon \sum_{k\neq j} p^2_{ik}).
\end{equation}
Using the above equations, along with equation \ref{TaylorP}, the dynamics can be written as:
\begin{equation}
p_{ij}(t+\delta t)=p_{ij}(t)+\epsilon (P_{C,i}-P_{D,i})p_{ij}(p_{ij}-\sum_k p^2_{ik}).
\end{equation}
Using continuum theory, for evolution of networks ~\cite{Barab99-2}, after long enough time-steps in the limit $N \rightarrow \infty$, $\frac{\epsilon}{\delta t}\rightarrow \frac{\partial e}{\partial t}(e=1)\equiv E(t)$, by defining $\eta_i:=P_{C,i}-P_{D,i}$ the dynamics becomes as:
\begin{equation}
\frac{\partial p_{ij}}{\partial t}=E(t)p_{ij}{\eta}_i(p_{ij}-M_{ii}).
\end{equation}
Where, $M$ is defined as $M=PP^T$. Inspired from diffusion mechanism, we can write the dynamics for the evolution of $\eta (N \times 1)$. Node $i$ is connected to $N-1$ nodes. At each time-step, if any walker is at node $i$, at the next time-step will transit to the other nodes, and new walkers transit to node $i$. For every node $j\neq i$ a constructive (destructive) walker concentration probability equal to $P_{C,j} (P_{D,j})$ transits to node $i$, with a probability equal to $p_{ji}$. Hence, the expectation value of $P_{C,i}$ and $P_{D,i}$, after each time-step evolves as the following recursive equations:
\begin{equation}
P_{C,i}(t+\delta t)=\sum_{j\neq i} p_{ji}(t)P_{C,j}(t),
\end{equation}
and
\begin{equation}
P_{D,i}(t+\delta t)=\sum_{j\neq i} p_{ji}(t)P_{D,j}(t).
\end{equation}
Since diagonal elements of transition probability matrix are zero, using continuum theory the dynamics of $\eta$ is as the following:

\begin{equation}
\frac{\partial {\eta}}{\partial t}=\frac{1}{\delta t}(P^T-I){\eta}.
\end{equation}

Where $\delta t$ is the time step between transitions, $P^T$ is the transpose of transition probability matrix and $I(N \times N)$ is the identity matrix. During this work, we have considered $E(t)$ constant and low walker approximation is assumed for all the simulations.
%%%%%%%%%%%%%%%%%%%%%%%%%%%%%%%%%%%%%%%%%%%%%%%%%%%%
\section{ Stationary states and stability}\label{AppB}
The stationary state of the network is when the expectation value of the random variables $p_{ij}$ does not evolve in time. If the stationary state happens to be unstable, after large enough time, the deviation of exact values obtained from simulation from expectation values (solution of equations \ref{PartialP} and \ref{PartialEta}), lead the state of the network to completely deviate from the stationary solutions. Since the elements of $P$ are the expectation values for transition probabilities, the $i$th column of probability matrix, vector $p^{(i)}=(p_{i1},...,p_{iN})^T$, satisfies the following conditions:

\begin {equation}\label{A1}
0\leq p_{ik} \leq 1, \forall k \in \{ 1 \ldots N \},
\end {equation}
\begin {equation}\label{A2}
\sum_l p_{il}=1.
\end {equation}

Element $p_{ij}$ is in stationary state if $\frac{\partial{p_{ij}}}{\partial{t}}=0$. Using equation \ref{PartialP}, this condition implies:

\begin {equation}\label{A3}
p_{ij}=\sum_k p^2_{ik},
\end{equation}

or

\begin{equation}
p_{ij}=0.
\end{equation}

Equation \ref{A3} can be written as $(p_{ij}-1/2)^2+\sum_{k \neq j}^N p^2_{ik}=(1/2)^2$. The set of all possible vectors, $p^{(i)}$, satisfying the third condition (equation \ref{A3}) is a $N$ dimensional hypersphere with radius equal to $\frac{1}{2}$, centered at $p_{i,j}=\frac{1}{2}, p_{i,k\neq j}=0$. The set of all vectors satisfying equation \ref{A2} is a $N$ dimensional hyperplane. Hence, the set of all vectors satisfying the three conditions (equations \ref{A1}, \ref{A2} and \ref{A3}) is the intersection of the hyperplane and hypersphere, satisfying $p^{(i)}\in [0,1]^N$, which is a $N-1$ dimensional hypersphere. The $N-1$ dimensional hypersphere corresponding to the stationary state of node $ij$ is expressed by $\Lambda^{ij}$.

The global stationary state for node $i$ is obtained when for every $k\neq i \in \{1 \ldots N\}$, $p_{ik}$ be in its stationary state. There are two global stationary states for the network:

a) If equation \ref{A3} is satisfied for all destination nodes, this global stationary state is the intersection of all $\Lambda^{ik}$ hyperspheres. If $p_{ik}=\sum_j p^2_{ij}$, be satisfied for all $k\neq i$, it reveals for all $k,l\neq i \in \{1 \ldots N\}$, $p_{ik}=p_{il}$. Hence, assuming the normalization condition this stationary state is $\forall k\neq i, p_{ik}=1/(N-1)$. This stationary state is called the uniform stationary state. The variation of equation \ref{PartialP} is as:

\begin{eqnarray}\label{A5}
\Delta \frac{\partial p_{ij}}{\partial t}&=&
(\frac{\epsilon}{\delta t})( \Delta (\eta_i p_{ij})(p_{ij}-\sum_k p^2_{ik})\\ \nonumber&+&\eta_i p_{ij}(\Delta p_{ij}- \sum_k (2p_{ik} \Delta p_{ik} ))).
\end{eqnarray}

Stationary state for node $i$ is stable if $\Delta \frac{\partial p_{ij}}{\partial t} \Delta p_{ij} \leq 0$ . For uniform stationary state, the first part of the right hand side of equation \ref{A5} is zero. The normalization condition implies $\sum_k \Delta p_{ik}=0$. In steady state if the walkers be distributed uniformly among nodes, $\Delta \frac{\partial p_{ij}}{\partial t}=(\frac{\epsilon}{\delta t})\eta_i \Delta (p_{ij})=(\frac{\epsilon}{\delta t})(2x-1) \Delta (p_{ij})$. Therefore node $i$ in steady state is stable when $x<0.5$.

b) The other stationary state for node $i$ is when there exist a node $j \neq i$, such that $p_{ij}=1$, and hence due to the normalization condition, for all $k \neq j, p_{ik}=0$. In this stationary state, which is called the isolated stationary state, the first part of the right hand side of equation \ref{A5} is zero, for the high weighted link ($p_{ij}=1$), and in steady state if the walkers are distributed uniformly among nodes, $\Delta \frac{\partial p_{ij}}{\partial t}=-(\frac{\epsilon}{\delta t})\eta_i \Delta (p_{ij})=-(\frac{\epsilon}{\delta t})(2x-1) \Delta (p_{ij})$. For the vanishing links $p_{ik}=0 (k \neq i, j$), the second part of the right hand side of equation \ref{A5} is zero, and by assuming the network in its steady state the variation is as $\Delta \frac{\partial p_{ij}}{\partial t}=-(\frac{\epsilon}{\delta t})\eta_i \Delta (p_{ij})=-(\frac{\epsilon}{\delta t})(2x-1) \Delta (p_{ij})$. This stationary state is stable if $x>0.5$. When $x=0.5$, the solution of equations \ref{PartialP} and \ref{PartialEta} is critical. The simulation results reveal that weight distributions are as power-law with scale-free properties, and even after long time-steps network doesn't reach steady state.
%%%%%%%%%%%%%%%%%%%%%%%%%%%%%%%%%%%%%%%%%%%%%%%%%%%%%%%%%%%%%%%%%%%%%%%%%%%%%%%%
%%%%%%%%%%%%%%%%%%%%%%%%%%%
\section*{References}
\providecommand{\newblock}{}

%%%%%%%%%%%%%%%%%%%%%%%%%%%%%%%%%%%%%%%%%%
\end{document}